\documentclass[journal]{IEEEtran}

\usepackage{pgfplots}
\usepackage[bookmarks,colorlinks]{hyperref}
\usepackage[linesnumbered,ruled,lined]{algorithm2e}
\usepackage{enumitem}
\usepackage{algpseudocode}
\usetikzlibrary{shapes.multipart,intersections}
\usepackage{cite}
\usepackage{amsmath,amssymb,amsfonts,amsthm,steinmetz}
\usepackage{graphicx}
\usepackage{mathrsfs}  
\usepackage{textcomp}
\usepackage{acronym}
\usepackage{xcolor}
\usepackage{upgreek,xspace}
\usepackage{array}
\usepackage{tikz}
\usetikzlibrary{calc}
\makeatletter
\newcommand{\gettikzxy}[3]{%
  \tikz@scan@one@point\pgfutil@firstofone#1\relax
  \edef#2{\the\pgf@x}%
  \edef#3{\the\pgf@y}%
}
\makeatother

\usepackage[draft]{todonotes}

\usepackage{esvect}

\usepackage{times}
\usepackage{bm}
\usepackage{amsmath}
\usepackage{amssymb}
\usepackage{stmaryrd}
\usepackage{babel}
\usepackage{graphics, graphicx}
\usepackage{xcolor}
\usepackage{gensymb}
\usepackage{cite}
\usepackage{enumitem}
\usepackage{url}

\begin{document}

\title{Experimental Multiport-Network Parameter Estimation and Optimization for Multi-Bit RIS}

\author{Philipp~del~Hougne,~\IEEEmembership{Member,~IEEE}
\thanks{This work was supported in part by the ANR France 2030 program (project ANR-22-PEFT-0005), the ANR PRCI program (project ANR-22-CE93-0010), the European Union's European Regional Development Fund, and the French Region of Brittany and Rennes M\'etropole through the contrats de plan \'Etat-R\'egion program (projects ``SOPHIE/STIC \& Ondes'' and ``CyMoCoD'').}
\thanks{
P.~del~Hougne is with Univ Rennes, CNRS, IETR - UMR 6164, F-35000, Rennes, France (e-mail: philipp.del-hougne@univ-rennes.fr).
}
}

\maketitle

\begin{abstract}
Physics-consistent \textit{theoretical} studies on RIS-parametrized wireless channels use  models from multiport-network theory (MNT) to capture mutual-coupling (MC) effects. However, \textit{in practice}, RIS design and radio environment are partially or completely unknown. We fill a research gap on how to estimate the MNT model parameters in such experimentally relevant scenarios. Our technique efficiently combines closed-form and gradient-descent steps, and it can be applied to multi-bit-programmable RIS elements. 
We discuss inevitable (but operationally irrelevant) parameter ambiguities. 
We experimentally validate our technique in an unknown rich-scattering environment parametrized by eight 6-bit-programmable RIS elements of unknown design. 
We experimentally evaluate the performance of RIS configurations optimized with the estimated MNT model and an MC-unaware cascaded model.
While the models differ in accuracy by up to 17~dB, the end-to-end performance differences are small.
\end{abstract}


\section{Introduction}

Tailoring wireless channels with reconfigurable intelligent surfaces (RISs) is a promising emerging technology for next-generation wireless networks. Physics-consistent models based on multiport-network theory (MNT) that capture the mutual coupling (MC) between RIS elements are currently gaining traction~\cite{matteo_universal}. Therein, the tunable lumped component of a RIS element is described as a virtual port terminated by a tunable load. A key open question that arises in moving toward experimental RIS-aided systems is how to obtain the MNT model parameters in practice.

Theoretical studies generate the model parameters in closed form, which limits the considered scenarios to analytically tractable ones -- typically thin-wire RIS elements and free-space radio environments~\cite{gradoni_EndtoEnd_2020}. Practical RIS elements are usually patches that are not analytically tractable. Moreover, only contrived discrete-dipole scattering environments are analytically tractable~\cite{PhysFad,mursia2023saris}. Meanwhile, attempts to add fading ad hoc do not account for the contribution of multi-path propagation to the MC between the RIS elements.

Full-wave numerical studies can extract all required MNT parameters in a single simulation, irrespective of the complexity of the RIS design and the scattering environment~\cite{tapie2023systematic}. However, the simulation requires detailed knowledge of the RIS design and the radio environment, neither of which is typically available in practice. 
Exceptions are well-controlled experiments with known RIS design in free-space radio environments, where  MNT-descriptions of the RIS obtained in full-wave simulations were successfully validated~\cite{zhang2022macromodeling,zheng2024mutual}; however, such well-controlled conditions are unlikely in practice.
The RIS design usually differs from the assumed one due to fabrication inaccuracies and component tolerances; it can also be completely unknown if the RIS prototype is based on an undisclosed, proprietary design. Moreover, the geometry and material composition of a real-world radio environment is only known approximately at best. Even if RIS design and radio environment were known perfectly, simulating the many-wavelengths-large domain can be (prohibitively) costly.

Meanwhile, the  unambiguous experimental estimation of all MNT parameters of an arbitrarily complex linear, passive, time-invariant device under test (DUT) with a subset of ports terminated by a specific tunable load network was recently achieved with ``Virtual VNA'' techniques~\cite{del2024virtual2p0,del2025virtual3p0,tapie2025scalable}. The latter have both closed-form and gradient-descent versions. 
In principle, treating the static parts of the radio environment (including structural scattering by antennas and RIS elements) as the DUT, these ``Virtual VNA'' techniques could hence unambiguously identify the MNT parameters without knowing the RIS elements' structural design nor the radio environment. However, the Virtual VNA techniques require that the load network's characteristics are known and include at least three distinct loads for each terminated DUT port, as well as coupled loads between adjacent DUT ports. The latter requirement is only satisfied by special ``beyond-diagonal'' RIS (BD-RIS)~\cite{tapie2025beyond}. Altogether, it is thus generally unlikely that the MNT parameters of a RIS-parametrized radio environment can be estimated unambiguously in practice. However, lifting all ambiguities is in fact not always required. To optimize a RIS for wireless communications functionalities, \textit{any set of parameters that correctly maps the RIS configuration to the corresponding end-to-end wireless channel matrix is sufficient}.

The first technique for identifying (an inevitably ambiguous) set of MNT parameters to describe an experiment involving an unknown 1-bit RIS design and unknown radio environment estimated all parameters at once via gradient descent~\cite{sol2024experimentally}.\footnote{\cite{sol2024experimentally} used a coupled-dipole formulation that is operationally equivalent to the scattering-parameter representation of MNT [Sec.~II.C.2,~\cite{del2025physics}].} Importantly, the number of parameters only depends on the number of transmitters, receivers and RIS elements~\cite{sol2024experimentally}; it is hence independent of the complexity of the RIS design and the complexity of the radio environments. However, the pure gradient-descent approach from~\cite{sol2024experimentally} does not take advantage of closed-form techniques that can robustly and efficiently estimate certain MNT parameters. For instance, a vector collinear with the one containing the transmission coefficients from the antennas to the RIS elements can be robustly retrieved via a singular value decomposition (SVD) of the difference between two channel matrix measurements~\cite{sol2024optimal}. Moreover, \cite{sol2024experimentally} only explored dictionary search for RIS optimization.

In this Letter, we fix these shortcomings. 
Our contributions are summarized as follows. 
\textit{First}, we present a robust MNT-parameter-estimation technique that efficiently combines closed-form and gradient-descent steps, and that can be applied to RIS with multi-bit element-wise programmability. 
\textit{Second}, we experimentally validate the technique in an unknown rich-scattering environment using the first RIS prototype with element-wise 6-bit programmability.\footnote{To the best of our knowledge, RIS prototypes with more than 2-bit element-wise programmability were not reported to date.}
\textit{Third}, we examine how the accuracy scales with the number of antennas and the number of measurements available for the gradient-descent step.
\textit{Fourth}, we conduct MC-aware RIS optimizations based on the estimated model parameters and validate the optimized configurations experimentally.
\textit{Fifth}, we benchmark model accuracy and end-to-end performance against two MC-unaware models.

\textbf{Remark:} Our technique is not limited to conventional RIS. It can be directly applied to BD-RISs and stacked intelligent metasurfaces (SIMs) whose hardware can be conceptually partitioned into a static part and a set of tunable lumped elements. Thus, the same MNT model applies in all cases.\footnote{There exists hence a physics-consistent \textit{diagonal} representation for BD-RIS that reveals the independently tunable degrees of freedom~\cite{del2025physics}. The entries of the conventional ``beyond-diagonal'' programmable matrix are correlated if realistic hardware implementations are considered~\cite{del2025physics,tapie2025beyond}.} This fact was already demonstrated by directly applying the algorithm for conventional RIS from \cite{sol2024experimentally} to a BD-RIS in~\cite{del2025physics}.

\textit{Notation:} 
$\mathbf{A} = \mathrm{diag}(\mathbf{a})$ denotes the diagonal matrix $\mathbf{A}$ whose diagonal entries are $\mathbf{a}$. 
$\mathbf{A}_\mathcal{BC}$ denotes the block of the matrix $\mathbf{A}$ whose row [column] indices are in the set $\mathcal{B}$ [$\mathcal{C}$]. 
$\mathcal{B}_i$ is the singleton containing the $i$th entry of $\mathcal{B}$.
$a_i$ denotes the $i$th entry of the vector $\mathbf{a}$. 
$a_{ij}$ denotes the $(i,j)$th entry of the matrix $\mathbf{A}$. 
$^\top$ and $^\dagger$ denote the transpose and transpose-conjugate operations, respectively. 
$\mathbf{1}_q$ denotes a $q\times 1$ vector whose entries are unity.

\section{System Model}
\label{sec_system_model}

As mentioned, we physics-consistently describe the RIS-parametrized radio environment in terms of MNT by partitioning it into static parts and a set of tunable lumped elements. A tunable lumped element admits a description as a virtual lumped port terminated by a tunable load.
We make the following three assumptions. \textit{First}, the static parts of the RIS-parametrized radio environment are linear, passive and reciprocal. \textit{Second}, the generators and detectors connected to the $N_\mathrm{T}$ transmitting and $N_\mathrm{R}$ receiving antennas are matched to a common reference impedance $Z_0$. \textit{Third}, the $N_\mathrm{A}=N_\mathrm{T}+N_\mathrm{R}$ antenna ports and the $N_\mathrm{S}$ tunable elements (each associated with one RIS element) are sufficiently small to admit a description as lumped components. Hence, we do \textit{not} make any assumptions about the RIS unit cell design (e.g., its structural scattering), nor about the radio environment's complexity (e.g., free space vs. rich scattering).

Under these general assumptions, the static parts of the RIS-parametrized radio environment constitute an $N$-port system characterized by the scattering matrix $\mathbf{S}\in\mathbb{C}^{N \times N}$, where $N=N_\mathrm{A}+N_\mathrm{S}$ and the same reference impedance $Z_0$ is assumed at all ports to define $\mathbf{S}$. $\mathbf{S}$ compactly accounts for all scattering phenomena between the ports, importantly without requiring explicit descriptions of structural or environmental scattering. 

We summarize the scattering properties of the $N_\mathrm{S}$ tunable lumped elements in a vector $\mathbf{c}\in\mathbb{C}^{N_\mathrm{S}}$ whose $i$th entry is the reflection coefficient of the individual load terminating the $i$th virtual port (associated with the $i$th RIS element).  

According to MNT~\cite{matteo_universal,del2025physics}, the end-to-end wireless channel $\mathbf{H}\in\mathbb{C}^{N_\mathrm{R}\times N_\mathrm{T}}$ is related to $\mathbf{S}$ and $\mathbf{\Phi}=\mathrm{diag}(\mathbf{c})$ via
\begin{equation}
    \mathbf{H} = {\mathbf{S}}_\mathcal{RT} +{\mathbf{S}}_\mathcal{RS} \left(\mathbf{\Phi}^{-1} - {\mathbf{S}}_\mathcal{SS}  \right)^{-1} {\mathbf{S}}_\mathcal{ST},
    \label{eq1}
\end{equation}
where $\mathcal{R}$, $\mathcal{T}$, and $\mathcal{S}$ denote the sets of port indices associated with receiving antennas, transmitting antennas, and RIS elements, respectively. Assuming  all entries of ${\mathbf{S}}_\mathcal{SS}$ vanish (i.e., zero MC between the RIS elements), (\ref{eq1}) specializes to the widespread MC-unaware, simplified, cascaded model:
\begin{equation}
    \mathbf{H}_\mathrm{casc} = {\mathbf{S}}_\mathcal{RT} +{\mathbf{S}}_\mathcal{RS} \mathbf{\Phi} {\mathbf{S}}_\mathcal{ST}.
    \label{eq2}
\end{equation}

\section{Problem Statement}

Practical RIS prototypes cannot realize arbitrary $\mathbf{c}$. Instead, the $i$th RIS element's virtual port can be terminated by one of $m$ possible loads that are summarized in the vector $\mathbf{s}_i\in\mathbb{C}^m$. For simplicity, and in line with most prototypes, we assume that the available loads are the same for each RIS element, i.e., $\mathbf{s}_i = \mathbf{s} \ \forall \ i$. 
We can relax this assumption to the case in which the virtual ports are terminated by a cascade of a static two-port system and a tunable load, where we only assume that the possible states of the tunable loads are identical while the static two-port systems may differ. This works because we can equivalently interpret these two-port systems as  part of the radio environment. 

The wireless practitioner reconfigures the RIS control sequence $\mathbf{w} \in \{1,2,\dots,m\}^{N_{\mathrm S}}
$ and measures the corresponding $\mathbf{H}$. 
The corresponding physically implemented reflection coefficients are $\mathbf{c}=f(\mathbf{w})$, where $f$ is a function that assigns the $w_i$th entry of $\mathbf{s}$ to the $i$th entry of $\mathbf{c}$.
However, neither $\mathbf{S}$ nor $\mathbf{s}$ are generally known, thwarting the use of the MNT model in (\ref{eq1}).

Our goal is to estimate a matching set of parameters $\{\tilde{\mathbf{S}}_\mathcal{RT},\tilde{\mathbf{S}}_\mathcal{RS},\tilde{\mathbf{S}}_\mathcal{SS},\tilde{\mathbf{S}}_\mathcal{ST},\tilde{\mathbf{s}}\}$ that we can use in lieu of the true set of parameters $\{{\mathbf{S}}_\mathcal{RT},{\mathbf{S}}_\mathcal{RS},{\mathbf{S}}_\mathcal{SS},{\mathbf{S}}_\mathcal{ST},{\mathbf{s}}\}$ to map any given $\mathbf{w}$ to its corresponding $\mathbf{H}$ using (\ref{eq1}). We hence do \textit{not} seek an unambiguous estimate of ${\mathbf{S}}$ and ${\mathbf{s}}$.

\section{Parameter Estimation}

\subsection{MC-Aware MNT Model}

Our approach is summarized as follows. \textit{First}, we fix $\tilde{\mathbf{S}}_\mathcal{RT}$ with a single measurement, without any calculations. \textit{Second}, we fix $\tilde{\mathbf{S}}_\mathcal{RS}$ and $\tilde{\mathbf{S}}_\mathcal{ST}^\top$ up to column-wise scaling factors using $N_\mathrm{S}$ additional measurements and SVDs. Third, we jointly fix the scaling factors, as well as $\tilde{\mathbf{S}}_\mathcal{SS}$ and $\tilde{\mathbf{s}}$, using $n$ additional measurements via gradient descent.

\textit{Step 1:} We define a reference control vector $\mathbf{w}_0$ and measure the corresponding $\mathbf{H}(\mathbf{w}_0)$. We define $\tilde{\mathbf{S}}_\mathcal{RT} \triangleq  \mathbf{H}(\mathbf{w}_0)$. $\mathbf{w}_0$ may be chosen arbitrarily; for simplicity, we use $\mathbf{w}_0 = \mathbf{1}_{N_\mathrm{S}}$. To be clear, we do \textit{not} assume that $\mathbf{w}_0$ corresponds to terminating the RIS elements with matched loads; hence, our estimated MNT parameters generally differ from the physical ones because $\tilde{s}_1$ should vanish if $\tilde{\mathbf{S}}_\mathcal{RT} \triangleq  \mathbf{H}(\mathbf{w}_0)$. 

\textit{Step 2:} We define a control vector $\mathbf{w}_i$ that equals $\mathbf{w}_0$ except for its $i$th entry. We measure the corresponding $\mathbf{H}(\mathbf{w}_i)$ and compute $\mathbf{\Delta}_i \triangleq  \mathbf{H}(\mathbf{w}_i)-\mathbf{H}(\mathbf{w}_0)$. Since $\mathbf{w}_i$ and $\mathbf{w}_0$ only differ by one entry, inspection of (\ref{eq1}) reveals that $\mathbf{\Delta}_i$ must be a rank-one matrix constructed by vectors that are colinear with $\tilde{S}_{\mathcal{RS}_i}$ and $\tilde{S}_{\mathcal{S}_i\mathcal{T}}$~\cite{sol2024optimal}. Specifically, the SVD yields $\mathbf{\Delta}_i=\mathbf{U}_i\mathbf{\Sigma}_i\mathbf{V}_i^\dagger$, where  $\mathbf{\Sigma}_i$ is a diagonal matrix containing the singular values (in descending order); the first column of $\mathbf{U}_i$ (resp. $\mathbf{V}_i$) is the first left (resp. right) singular vector of $\mathbf{\Delta}_i$, which we denote by $\mathbf{u}_i$ (resp. $\mathbf{v}_i$). Hence, we define $\tilde{S}_{\mathcal{RS}_i} \triangleq a_i \mathbf{u}_i$ and $\tilde{S}_{\mathcal{S}_i\mathcal{T}} \triangleq b_i \mathbf{v}_i^\dagger$, where $a_i$ and $b_i$ are scaling factors that remain to be determined. We repeat this procedure for each RIS element in turn. The $i$th entry of $\mathbf{w}_i$ can be chosen arbitrarily as long as it differs from the $i$th entry of $\mathbf{w}_0$; choices yielding larger norms of $\mathbf{\Delta}_i$ are preferable in the presence of noise. 

\textit{Step 3:} We define a set of $n$ random realizations of $\mathbf{w}$ and measure the corresponding realizations of $\mathbf{H}$. We then perform gradient descent to jointly estimate $\mathbf{a}=[a_1, \dots, a_{N_\mathrm{S}}]$, $\mathbf{b}=[b_1, \dots, b_{N_\mathrm{S}}]$, $\tilde{\mathbf{S}}_\mathcal{SS}$, and $\tilde{\mathbf{s}}$; we impose that $\tilde{\mathbf{S}}_\mathcal{SS}$ is symmetric due to reciprocity. The total number of complex-valued unknowns is $2N_\mathrm{S} + N_\mathrm{S}(N_\mathrm{S}+1)/2 + m$. The cost to be minimized is the average magnitude of the difference between the entries of the measured and predicted $\mathbf{H}$. 

Compared to~\cite{sol2024experimentally}, Steps 1 and 2 require only $1+N_\mathrm{S}$ measurements and reduce the number of complex-valued parameters to be estimated by gradient descent by $N_\mathrm{R}N_\mathrm{T} + (N_\mathrm{R}+N_\mathrm{T}-2)N_\mathrm{S}$, where the first term corresponds to $\tilde{\mathbf{S}}_\mathcal{RT}$ and the second term corresponds to $\tilde{\mathbf{S}}_\mathcal{RS}$ and $\tilde{\mathbf{S}}_\mathcal{ST}$ while accounting for $\mathbf{a}$ and $\mathbf{b}$. Step 1 is trivial and Step 2 is known to be experimentally robust from~\cite{sol2024optimal}, as well as from challenging Virtual VNA experiments~\cite{tapie2025scalable} (although we do not seek to lift ambiguities here). The reasons for estimating $\tilde{\mathbf{S}}_\mathcal{SS}$ via gradient descent rather than in closed form in Step 3 are two-fold. On the one hand, corresponding estimates of $\tilde{\mathbf{S}}_\mathcal{SS}$ in closed-form Virtual VNA methods are notably more vulnerable to noise than the gradient-descent Virtual VNA counterparts~\cite{tapie2025scalable}. On the other hand, $\mathbf{s}$ is unknown here (whereas it is known for the Virtual VNA), which might preclude closed-form techniques to estimate $\tilde{\mathbf{S}}_\mathcal{SS}$ for $m>2$. Indeed, for $m>2$, one cannot arbitrarily fix the entries $\tilde{\mathbf{s}}$ because this would cause deviations from the relations between the entries of $\mathbf{s}$ that could not be compensated elsewhere. The case of $m=2$ is an exception in that regard. 

The total number of measurements is $1+N_\mathrm{S}+n$.

\subsection{MC-Unaware Benchmark Models}
 
\textit{CASC:} We proceed as for MNT, except that we fix all entries of $\tilde{\mathbf{S}}_\mathcal{SS}$ to zero (i.e., we assume that there is no MC), such that (\ref{eq1}) collapses to (\ref{eq2}).

\textit{LR:} Based on the available $1+N_\mathrm{S}+n$ measurements, we perform a multivariable linear regression to find the best affine mapping from $\mathbf{w}$ to the $(i,j)$th channel coefficient: $h_{ij}^\mathrm{LR} = \mathbf{p}^\top\mathbf{w}+q$, where $\mathbf{p}\in\mathbb{C}^{N_\mathrm{S}}$ and $q\in\mathbb{C}$. 
LR captures neither the MC-induced non-linearity in the mapping from $\mathbf{c}$ to $\mathbf{H}$ nor the non-linearity in the mapping from $\mathbf{w}$ to $\mathbf{c}$. The latter is only relevant for $m>2.$

\section{Experimental Validation}

\subsection{Experimental Setup}

\begin{figure}[b]
\centering
\includegraphics[width=\columnwidth]{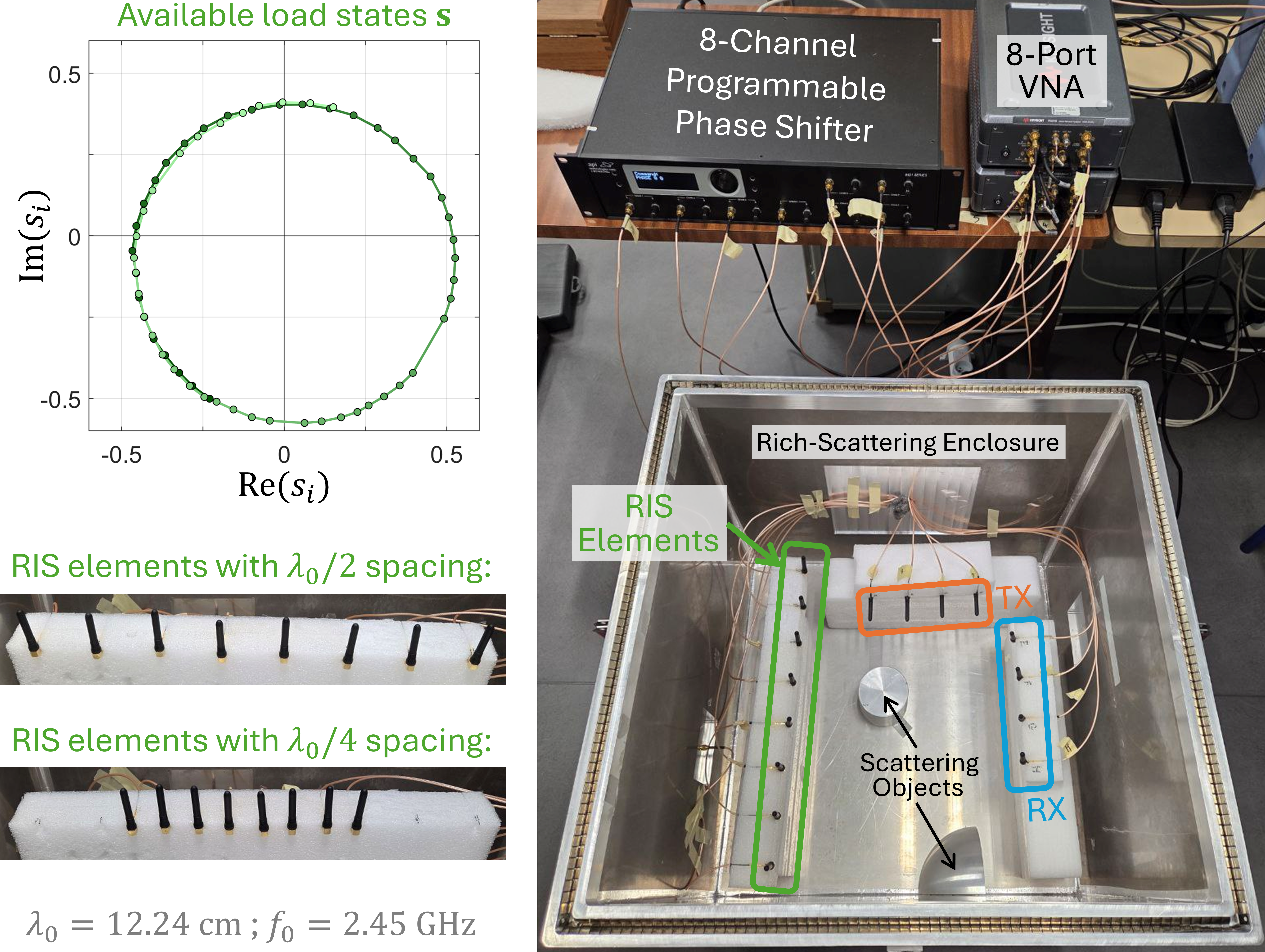}
\caption{Experimental setup (top cover removed to show interior).}
\label{fig1}
\end{figure}

Our RIS prototype with element-wise 6-bit programmability is built by connecting a set of eight commercial WiFi antennas (ANT-24G-HL90-SMA) to an eight-channel reconfigurable phase shifter (API Weinschel, Inc. 8421-A1-08-FS) via coaxial cables (RG316). Each phase shifter channel can be independently configured to any of $2^6=64$ states (their reflection coefficients are plotted in Fig.~\ref{fig1}); all channels are nominally identical. Our modular RIS design enables adjustments of the RIS element spacing (and thereby the MC strength) and direct measurements of $\mathbf{s}$ (not used for parameter estimation). 

Our transmit and receive arrays each comprise four parallel, half-wavelength-spaced antennas (ANT-24G-HL90-SMA). 
As seen in Fig.~\ref{fig1}, they are placed alongside our RIS inside a reverberation chamber (59~$\times$~60~$\times$~58~$\mathrm{cm}^3$) which yields a rich-scattering environment that precludes a priori channel knowledge. The transmitting and receiving antennas are oriented perpendicular to each other to minimize the direct line-of-sight channel.
We use an eight-port VNA (two cascaded Keysight P5024B 4-port VNAs) to measure $\mathbf{H}$ at our operating frequency of 2.45~GHz. Note that, despite our modular RIS design, directly measuring $\mathbf{S}$ is challenging because it would require a 16-port VNA. 

We conduct our experiment under well-controlled conditions: The signal-to-noise ratio (SNR) is 65.9~dB (estimated based on repeated measurements in quick succession of $\mathbf{H}$ for the same $\mathbf{w}$) and the stability is 53.2~dB (estimated like SNR, but based on measurements taken intermittently over the course of our experiment).

\subsection{Experimental Parameter Estimation}

We quantify a model's accuracy with a metric that is defined similar to an SNR, treating the model error as ``noise'':
\begin{equation}
    \zeta = \frac{\mathrm{SD}\left[h_{ij}^\mathrm{GT}(\mathbf{w})\right]}{\mathrm{SD}\left[h_{ij}^\mathrm{GT}(\mathbf{w}) - h_{ij}^\mathrm{PRED}(\mathbf{w})\right]},
    \label{eq_zeta}
\end{equation}
where $\mathrm{SD}$ denotes the standard deviation across all entries of $\mathbf{H}$ and 3000 random, unseen realizations of $\mathbf{w}$; the superscripts denote the experimentally measured ground truth (GT) and the model's prediction (PRED).

\begin{figure}
\centering
\includegraphics[width=0.7\columnwidth]{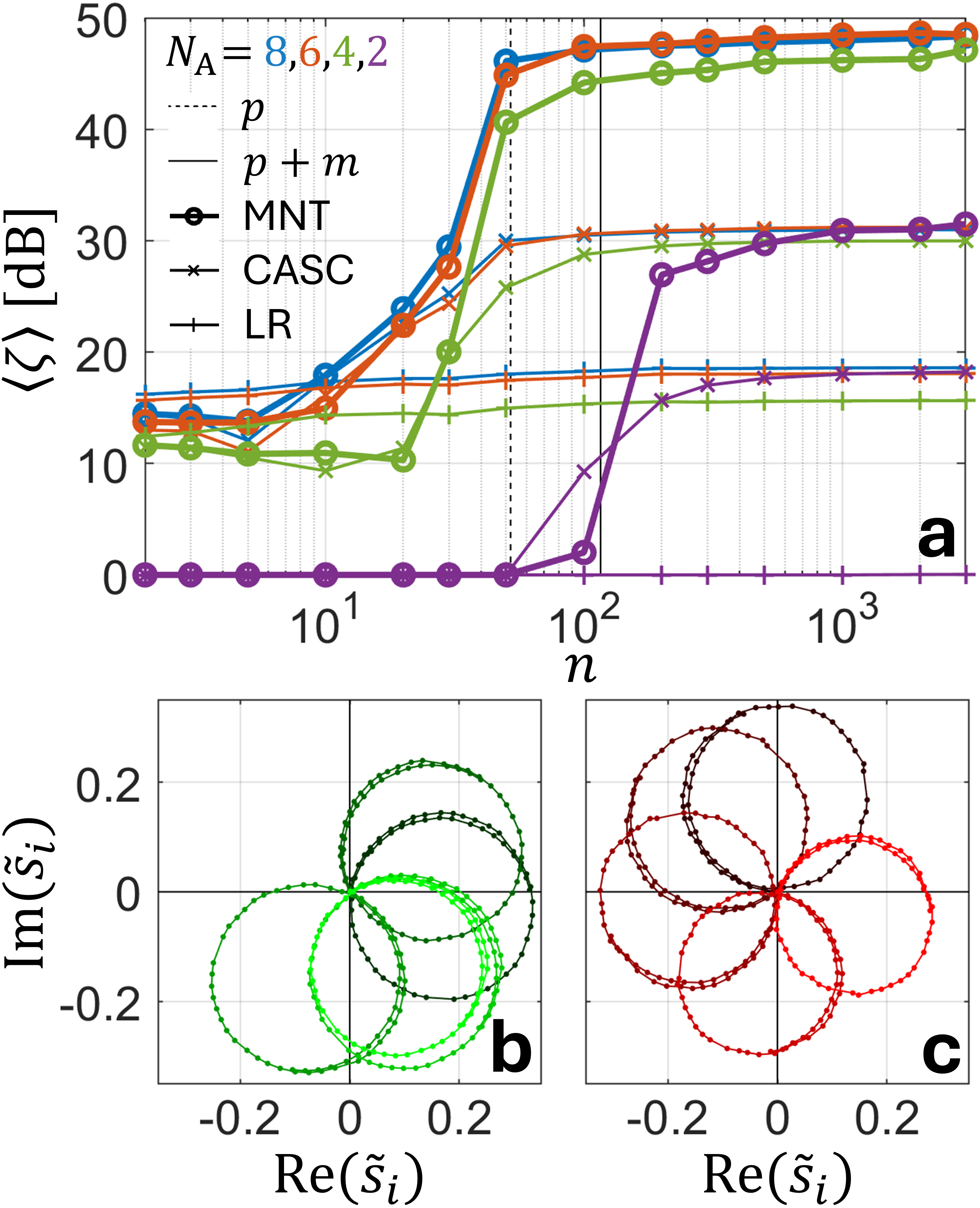}
\caption{(a) Average accuracy achieved by the MNT~(\texttt{o}), CASC~($\times$), and LR~(\texttt{+}) models as a function of $n$ for four different values of $N_\mathrm{A}$ (color-coded); $p=2N_\mathrm{S}+N_\mathrm{S}(N_\mathrm{S}+1)/2$; $\langle\zeta\rangle$ is floored at 0~dB. (b,c) Estimated values of $\tilde{\mathbf{s}}$ with MNT (b) and CASC (c) with five distinct random seeds.}
\label{fig2}
\end{figure}

We present a systematic analysis of how $\zeta$ depends on $N_\mathrm{A}$ and $n$ in Fig.~\ref{fig2}a for $\lambda_0/2$ spacing between RIS elements; we observe the same trends with $\lambda_0/4$ spacing. Each parameter estimation is repeated five times with different random seeds for the gradient descent in Step 3; seed-dependent fluctuations of $\zeta$ are minimal.
For simplicity, we set $N_\mathrm{T}=N_\mathrm{R}=N_\mathrm{A}/2$. 

We observe that a larger value of $N_\mathrm{A}$ generally yields a higher model accuracy. This makes sense because each measurement contains more information. However, the marginal benefit of increasing $N_\mathrm{A}$ appears to diminish rapidly. Whereas the benefit of going from $N_\mathrm{A}=2$ to $N_\mathrm{A}=4$ is substantial (increase of $\langle\zeta\rangle$ by 15.6~dB for the MNT model and by 11.8~dB for the CASC model, for $n=3000$), going to $N_\mathrm{A}=6$ yields only small additional benefits (MNT: 1.4~dB, CASC: 1.2~dB, for $n=3000$), and going further to $N_\mathrm{A}=8$ yields essentially no further benefits.
We also observe that a larger value of $n$ generally yields a higher model accuracy, although this trend saturates when $n$ is increased beyond the order of magnitude of the number of unknowns (indicated by a vertical line).
The CASC model displays the same qualitative behaviors as the MNT model in terms of its accuracy, but its accuracy is roughly 17~dB lower for large values of $n$. Specifically, the highest achieved MNT accuracy is 48.5~dB whereas the corresponding CASC accuracy is 31.2~dB. 
Meanwhile, the LR accuracy remains always below 20~dB and only weakly scales with $n$. For $N_\mathrm{A}=2$, the LR accuracy never exceeds 0~dB.

We also visualize in Fig.~\ref{fig2}b,c the MNT and CASC estimates of the available loads, i.e., $\tilde{\mathbf{s}}$, obtained with five different seeds at $n=3000$. Upon visual inspection, their resemblance with $\mathbf{s}$ from Fig.~\ref{fig1} is apparent. Specifically, every $\tilde{\mathbf{s}}$ appears to be the result of some scaling, rotation, and translation of $\mathbf{s}$. Different random seeds yield roughly (but not exactly) the same scaling but significantly different rotations, as seen in Fig.~\ref{fig2}b,c. It is also apparent that all plots of $\tilde{\mathbf{s}}$ cross at the origin, as expected because $\tilde{s}_1$ should vanish, as explained in Step 1.

\section{Performance Evaluation}

We now check to what extent the substantial accuracy differences between the three  models manifest themselves in terms of end-to-end communications performance.

\subsection{Key performance indicators (KPIs)}

\textit{SISO KPI 1:} We aim to maximize the channel gain $|h_{ij}|^2$. This is a commonly considered KPI in theoretical papers on RIS. Albeit on the antenna level, this KPI directly correlates with numerous end-to-end SISO communications metrics (e.g., capacity, bit-error rate, throughput).

\textit{SISO KPI 2:} We assume that the $i$th TX communicates with the $i$th RX, such that the signals from the $i$th TX to the $j$th RX, where $j\neq i$, constitute undesired interferences. 
The rate achieved by the $i$th TX is hence $R_i=\mathrm{log}_2\left( 1 + P_\mathrm{T}|h_{ii}|^2 / (\sum_{j\neq i}  P_\mathrm{T}|h_{ij}|^2 + \sigma^2) \right)$, where we assume that all TX transmit with power $P_\mathrm{T}$; $\sigma$ quantifies the noise strength. This scenario represents a SISO interference channel. Our KPI is $R_i$, assuming $P_\mathrm{T}/\sigma^2=100$~dB.

\textit{MIMO KPI 1:} We aim to maximize the capacity $C$ at low signal-to-noise ratio (SNR) by dominant eigenmode transmission. In this regime, $C=\mathrm{log}_2\left( 1+ P_\mathrm{T}\|\mathbf{H}\|^2/\sigma^2 \right)$ so that maximizing $C$ is equivalent to maximizing the spectral norm of the channel matrix: $\|\mathbf{H}\|^2$. Hence, our KPI is $\|\mathbf{H}\|^2$.

\textit{MIMO KPI 2:}  We aim to maximize the capacity $C$ at high SNR by multi-eigenmode transmission with uniform power allocation. In this regime,  $C=\mathrm{log}_2\left( \mathrm{det}\left( \mathbf{I}_{N_\mathrm{R}}+ P_\mathrm{T}\mathbf{H}\mathbf{H}^H/\sigma^2 \right)\right)$. Hence, our KPI is $C$, assuming $P_\mathrm{T}/\sigma^2=100$~dB.

\subsection{RIS Optimization}

For all three model-based optimizations, we use the same coordinate-descent algorithm initialized with the best RIS configuration among the $1+N_\mathrm{S}+n$ measured ones. For one RIS element after another, we test all possible 64 states and retain the one that maximizes our KPI. We keep looping over the RIS elements until the KPI has not improved during $N_\mathrm{S}$ iterations, indicating convergence of the coordinate descent. We efficiently conduct forward evaluations of the MNT model from (\ref{eq1}) using the Woodbury identity~\cite{prod2023efficient}. Explorations of more advanced gradient-based optimization algorithms are left for future work since we focus mainly on the MNT parameter estimation in this Letter.

\subsection{Experimental Results}

We benchmark our ability to maximize the four KPIs based on the three considered models against the average of the KPIs over 3000 random RIS configurations (denoted by \textit{RAND}).
We measure $\mathbf{H}$ for each optimized RIS configuration directly in our experiment to prevent any potential model-reality mismatch from influencing the reliability of the reported KPI values. For conciseness, we limit the presented results to a representative case with $N_\mathrm{T}=N_\mathrm{R}=4$, $n=3000$, and a RIS with $\lambda_0/2$ inter-element spacing. Our results with $\lambda_0/4$ spacing are very similar.

The achieved KPIs are summarized in Table~\ref{table_half}. The RIS-enabled performance enhancement appears to substantially depend on the KPI. Optimizing the RIS with our highly-accurate MNT model enables 67.5~\% and 86.5~\% performance enhancements over the RAND benchmark for the SISO KPIs 1 and 2, respectively. Using the less accurate CASC model instead causes only small performance degradations (achieving 64.8~\% and 82.8~\% performance enhancement, respectively). This is an important observation for wireless practitioners who may hence prefer the much lighter CASC parameter estimation. Meanwhile, the LR model perform rather poorly (21.0~\% and 12.5~\%). We observe that the MIMO KPIs are generally less affected by the RIS configuration; especially MIMO KPI 2 barely depends on how we configure our RIS. Compared to the RAND benchmark, we achieve performance enhancements of 15.4~\% and 2.0~\% for MIMO KPIs 1 and 2, respectively, based on our highly accurate MNT model. Again, the CASC model performs almost as well (14.5~\% and 1.9~\%) and the LR model performs poorly (7.0~\% and 0.8~\%). Altogether, these results raise doubts about the importance (in terms of end-to-end communications performance) of capturing MC in system models of RIS-parametrized channels.

\begin{table}[h]
\centering
\caption{KPIs for RIS with $\lambda_0/2$ spacing. SISO KPIs are averaged over those achieved with all possible choices of $i$ and $j$.}
\begin{tabular}{ 
    |p{2.3cm}|| 
    >{\centering\arraybackslash}p{1.0cm}|
    >{\centering\arraybackslash}p{1.0cm}|
    >{\centering\arraybackslash}p{1.0cm}|
    >{\centering\arraybackslash}p{1.0cm}|
}
 \hline
 KPI \textbackslash{} Method & RAND & MNT & CASC & LR \\
 \hline\hline
 $|h_{ij}|^2 \! \times \! 10^3$ &  11.7 &  19.6 &  19.3 &  14.2 \\
 $R_i  \times  10^2$ [bits/s/Hz] &  13.7 &  25.5 &  25.0 &  15.4 \\
 $\|\mathbf{H}\|^2 \! \times \! 10^2$ &  11.8 &  13.6 &  13.5 &  12.7 \\
 $C$ [bits/s/Hz]  &  109.2 &  111.3 &  111.3 &  110.1 \\
 \hline
\end{tabular}
\label{table_half}
\end{table}

\section{Conclusion}

We developed and experimentally validated the first hybrid MNT parameter estimation technique that efficiently combines closed-form and gradient-descent steps, for RIS with element-wise multi-bit programmability and without any knowledge of the RIS design or the radio environment.
Surprisingly, we observed that while a benchmark MC-unaware cascaded model's accuracy remained 17~dB below the MNT model's accuracy, it achieved almost the same performance enhancements in terms of end-to-end wireless communications metrics. This experimental evidence can give wireless practitioners confidence that MC-unaware cascaded models, which are widely used in theoretical studies, can yield very good experimental performances. Meanwhile, the developed MC-aware parameter estimation technique may find applications for reconfigurable wave systems featuring strong MC (e.g., BD-RISs, SIMs, dynamic metasurface antennas, wave-domain physical neural networks).

Looking forward, Step 2 can be refined using a sequence of RIS configurations in which multiple RIS element states differ from the reference configuration, with distinct but partially overlapping groups of differing RIS elements in subsequent configurations. 
In addition, a blockwise estimation of $\tilde{\mathbf{S}}_\mathcal{SS}$ and a gradient-based RIS optimization can facilitate the scaling to larger $N_\mathrm{S}$. Moreover, exploiting spectral correlations can ease the burden of broadband scenarios.

\bibliographystyle{IEEEtran}

% Generated by IEEEtran.bst, version: 1.14 (2015/08/26)
\providecommand{\noopsort}[1]{}\providecommand{\singleletter}[1]{#1}%

\end{document}